\documentclass{aa}

\usepackage{graphicx}
\usepackage{appendix}
\usepackage{subcaption}

\usepackage{txfonts}

\usepackage{color}
\usepackage{natbib,twoopt}
\usepackage[breaklinks]{hyperref}      
\bibpunct{(}{)}{;}{a}{}{,}             
\hypersetup{
  colorlinks,
  citecolor=blue,
  linkcolor=blue,
  urlcolor=blue,
}
\makeatletter
  \newcommandtwoopt{\citeads}[3][][]{\href{http://adsabs.harvard.edu/abs/#3}%
    {\def\hyper@linkstart##1##2{}%
     \let\hyper@linkend\@empty\citealp[#1][#2]{#3}}}
  \newcommandtwoopt{\citepads}[3][][]{\href{http://adsabs.harvard.edu/abs/#3}%
    {\def\hyper@linkstart##1##2{}%
     \let\hyper@linkend\@empty\citep[#1][#2]{#3}}}
  \newcommandtwoopt{\citetads}[3][][]{\href{http://adsabs.harvard.edu/abs/#3}%
    {\def\hyper@linkstart##1##2{}%
     \let\hyper@linkend\@empty\citet[#1][#2]{#3}}}
  \newcommandtwoopt{\citeyearads}[3][][]%
    {\href{http://adsabs.harvard.edu/abs/#3}
    {\def\hyper@linkstart##1##2{}%
     \let\hyper@linkend\@empty\citeyear[#1][#2]{#3}}}
\makeatother
\definecolor{cobalt}{rgb}{0.06, 0.2, 0.65}

\definecolor{smalt(darkpowderblue)}{rgb}{0.0, 0.2, 0.6}
\definecolor{forestgreen(traditional)}{rgb}{0.0, 0.5, 0.0}

\newcommand{\Msun}{M$_{\odot}$}
\newcommand{\Mjup}{M$_{\mathrm{J}}$}

\newcommand{\gppr}{\stackrel{>}{\scriptstyle \sim}}
\newcommand{\gappr}{\raisebox{-0.4ex}{$\gppr$}}
\newcommand{\lppr}{\stackrel{<}{\scriptstyle \sim}}
\newcommand{\lappr}{\raisebox{-0.4ex}{$\lppr$}}

\defcitealias{Villaver_2009}{VL09}
\defcitealias{rasioetal96-1}{R96}
\usepackage[normalem]{ulem}

\begin{document}

   \title{Predicted incidence of Jupiter-like planets around white dwarfs}
  
   \titlerunning{Predicted incidence of Jupiter-like planets around white dwarfs}

   \author{Alex Mauch-Soriano\inst{1,2},
          Matthias R. Schreiber
          \inst{1}
          \and
          Diego Correa\inst{1}
          \and
          Julio Pinilla
          \inst{1}
          \and
          Catalina Riveros-Jara\inst{1}
          \and
          Javiera Vivanco
          \inst{1}
          \and
          Maria Paula Ronco
          \inst{3}
          \and
          Diogo Belloni 
          \inst{4}
          \and
          Felipe Lagos-Vilches\inst{1}
          \and
          Wolfgang Brandner\inst{5}
          }

   \institute{Departamento de F\'isica, Universidad T\'ecnica Federico Santa Mar\'ia, Av. Espa\~na 1680, Valpara\'iso, Chile\\
              \email{alex.mauch@usm.cl}
         \and
         {Instituto de F\'isica, Pontificia Universidad Cat\'olica de Valpara\'iso, Casilla 4059, Valpara\'iso, Chile}
            \and
            {Instituto de Astrof\'{\i}sica de La Plata, CCT La Plata-CONICET-UNLP, Paseo del Bosque S/N (1900), La Plata, Argentina}
                \and
                 S\~ao Paulo State University (UNESP), School of Engineering and Sciences, Guaratinguetá, Brazil
                \and
                Max Planck Institute for Astronomy, Königstuhl 17, 69117 Heidelberg, Germany
                }
   \date{Received ; accepted}
  \abstract
   {Gas-giant planets and brown dwarfs have been discovered in large numbers around main-sequence stars and even evolved stars. In contrast, and despite ongoing imaging surveys using state-of-the-art facilities, only a handful of substellar companions to white dwarfs are known. It remains unclear whether this paucity reflects observational challenges or the consequences of stellar evolution.}
   {We aim to carry out population synthesis of substellar objects around white dwarfs to predict the fraction and properties of white dwarfs hosting substellar companions.}
   {We generated a representative population of white-dwarf progenitors (up to $4$\,\Msun) with substellar companions, adopting companion distributions derived from radial-velocity surveys of giant stars and a global age-metallicity relation. We then combined the stellar-evolution codes Modules for Experiments in Stellar Astrophysics (MESA) and Single Star Evolution (SSE) with standard prescriptions for mass loss and stellar tides to predict the resulting population of white dwarfs and their substellar companions.}
   {We find that the predicted fraction of white dwarfs hosting substellar companions in the Milky Way is, independent of uncertainties related to initial distributions, stellar tides, or stellar mass loss during the asymptotic giant branch, below $\sim3\pm1.5$ \%. The occurrence rate peaks at relatively low-mass ($\sim 0.53M_\odot$ to $\sim 0.66M_\odot$) white dwarfs and relatively young ($\sim 1$-$6$ Gyr) systems, where it can reach $\gappr3$\,\%. 
   The semimajor axes of the surviving companions range from
   $3-24$\,au with a median of $11$\,au. We estimate that $\sim95$\,\% of the predicted companions are gas-giant planets, which translates to a predicted general Jupiter-like planet occurrence rate around white dwarfs below $\sim2.9\pm1.4$\,\%. These occurrence rates might slightly increase if multi-planetary systems are considered. 
   Furthermore, owing to the strong dependence
   of companion occurrence on the metallicity of the white dwarf progenitor, the assumed age-metallicity relation strongly affects the predictions. Based on recent estimates of the local age-metallicity relation, we estimate that the fraction of white dwarfs with companions close to the Sun might reach $\lappr8$\,\%. }
   {If the planetary and brown dwarf companion distributions derived from intermediate-mass giant stars through radial velocity surveys reflect the characteristics of the true population, less than $3\,\pm1.5$\,\% of white dwarfs host substellar companions. Depending somewhat on the age-metallicity relation, this most likely represents an upper limit on possible detections because a significant number of companions might not be detectable with current facilities.}


   \maketitle
%

\section{Introduction}

Several thousand exoplanets are confirmed to exist around main-sequence stars,\footnote{\url{https://exoplanetarchive.ipac.caltech.edu/}} and on average each star in the Milky Way probably hosts more than one planet \citep{cassanetal12-1}. 
Virtually all known planet host stars will end their lives as white dwarfs (WDs)
One might therefore expect large numbers of exoplanets to exist around WDs. 

Although in specific cases the gravitational interactions between the planets can affect survival \citep{roncoetal20-1}, in general planets with orbits beyond a few astronomical units should mostly survive unscathed, while planets orbiting at separations closer than $\sim1$\,au should be engulfed during the giant phases of the host star \citep[e.g.,][]{villaver+livio07-1,veras16-1,Nordhaus-13}. 
Companions that survive the giant-star phases of their host stars are expected to migrate outward due to stellar mass loss. 

Several surveys of WDs 
using various direct imaging techniques to find surviving massive giant planets have been performed without a clear detection \citep{burleighetal02-1,debesetal05-1,hoganetal09-1,brandneretal21-1}. These results from high-contrast imaging campaigns are complemented by transit searches using Kepler data \citep{van-sluijsetal18-1} and surveys of WDs with infrared excess using Spitzer photometry \citep{farihietal08-1}. 

Despite these efforts, very few substellar companions to WDs have been detected so far. 
Two of the detected substellar objects, the hypothetical Neptune-like planet around WD\,J0914+1914 \citep{gaensickeetal19-1} and the transiting Jupiter-like planet around WD\,1856+534 \citep{vandenburgetal20-1}, orbit their central WDs with short orbital periods, which implies that they either must have been affected by other companions, or that they perhaps survived a common envelope phase \citep{munoz+petrovic20-1,lagosetal21-1,maldonadoetal21-1, Chamandy-21,Stephan-21}. 
In contrast, the $1.4$ Jupiter-mass planet discovered by \citet{blackmanetal21-1} orbits the central WD at a projected separation of $2.8\pm0.5$\,au, which suggests that it might have been affected by the metamorphosis of its host star mostly through orbital expansion as a consequence of mass loss. The origin of the very distant ($2500$\,au projected separation) substellar companion candidate to WD\,0806-661 \citep{luhmanetal12-1} and   
the rocky planet candidates potentially orbiting close binaries at large separations \citep{thorsettetal93-1,sirgurdssonetal03-1,luhmanetal12-1,zhangetal24-1} remains unclear. 
\citet{brandneretal21-1} argued that the low number of detected substellar objects around WDs is related to instrumental limitations. This situation might be changing as the James Webb Space Telescope (JWST) and new ground-based, high-contrast imaging instruments (e.g., ERIS at the VLT) have become available. 
Indeed, three candidate giant planets around WDs have recently been detected using the JWST \citep{limbachetal24-1,mullallyetal24-1}, and complementary surveys with ERIS are underway. Furthermore, in the near future large telescopes such as the Extremely Large Telescope (ELT) will probably revolutionize the field. 

While the prospect of providing a more robust observational census of planets orbiting WDs in the near future is promising, few theoretical predictions have been made. To the best of our knowledge, the only population-synthesis approach to predicting the occurrence rate of planets around WDs 
was published by \citet{andryushin+popov21-1}. These authors assumed a rather generic planet population around WD progenitor stars and focused on predicting the fractions of planets that are engulfed or ejected or remain in orbit despite the metamorphosis of their host star. While this in general represents a useful exercise, it does not allow us to predict the relative number of WDs that is expected to host a planet and how this fraction depends on WD parameters such as age (effective temperature) and mass. In addition, by using a generic planet population, \citet{andryushin+popov21-1} did not take into account possible dependencies of the planet-occurrence rate on stellar mass and metallicity. 
Therefore, despite clearly being interesting, the results obtained by \citet{andryushin+popov21-1} do not permit direct comparison with the results of ongoing observational surveys dedicated to planets around WDs. 

We followed a different population-synthesis approach to fill this gap. We generated a representative population of WD progenitors based on an age-metallicity relation and the initial mass function and assigned substellar companions with masses from $0.8$ up to $30$\,\Mjup~using occurrence rates derived from radial-velocity surveys of giant stars.
We then simulated the evolution of the host stars and the planetary orbits and predict that the general fraction of white dwarfs hosting a substellar companion is below $3\pm1.5$\,\% and that the vast majority of these companions are Jupiter-like gas-giant planets. 
This result is obtained assuming single planet systems. Taking into account additional companions that potentially expel part of the stellar envelope during the giant phases, this fraction could increase to $\lappr4.4\pm2.4$\%.   
Furthermore, the predictions of our population synthesis significantly depend on the assumed age-metallicity relation.
In what follows, we describe our model and present its main predictions, beginning with a description of the numerical codes used.

\section{Survival of substellar objects}

To determine the impact of stellar evolution on the orbits of substellar companions, we combined the 
Modules for Experiments in Stellar Astrophysics 
(MESA) code \citep[e.g.,][]{Jermyn2023} 
with an algorithm solving the equations describing the time evolution of the orbit of companions affected by stellar mass loss and stellar tides. This algorithm is called 
Final exoplAnet orbiTal dEstination around evolved Stars (FATES) and was developed by \citet{roncoetal20-1}. 
Both codes are described in detail in what follows. 

\subsection{Stellar evolution with MESA}

The MESA code (we used version r-23.5.1) allows users to reproduce the details of stellar evolution, albeit at high computational cost. Consequently, we chose to employ MESA exclusively for stars that host a planet. 

The MESA equation of state is a blend of the OPAL \citep{Rogers2002}, SCVH
\citep{Saumon1995}, FreeEOS \citep{Irwin2004}, HELM \citep{Timmes2000},
PC \citep{Potekhin2010}, and Skye \citep{Jermyn2021} EOSes. Radiative opacities are primarily from OPAL \citep{Iglesias1993,
Iglesias1996}, with low-temperature data from \citet{Ferguson2005}
and the high-temperature, Compton-scattering-dominated regime by
\citet{Poutanen2017}. Electron conduction opacities are from
\citet{Cassisi2007} and \citet{Blouin2020}. 
Nuclear reaction rates are from JINA REACLIB \citep{Cyburt2010}, NACRE \citep{Angulo1999} and
additional tabulated weak reaction rates \citep{Fuller1985, Oda1994,
Langanke2000}. Screening is included via the prescription of \citet{Chugunov2007}.
Thermal neutrino-loss rates are from \citet{Itoh1996}.

Additional key parameters in MESA for the survival of substellar companions are the mass-loss parameters. During the red giant branch (RGB), we assumed the model proposed by \citet{Reimers1975}, 
and for stars on the asymptotic giant branch (AGB) we used the relation suggested by \citet{Bloecker95-1}. 

Regarding convective overshooting, in principle it should always be included when there is convection. However, while constraints on core overshooting have been derived from asteroseismology, overshooting in the convective envelope remains poorly constrained. Therefore, we only included overshooting for convective cores following \citet{Schaller1992} and \citet{Herwig2000}.

\subsection{Orbital evolution of the companions}

To model the orbital evolution of the companions over the lifetime of their host star, we used the \textit{FATES} code developed by \citet{roncoetal20-1}, a Runge-Kutta-Fehlberg algorithm that integrates the equations 
describing the effects of mass loss and tidal forces on the orbit (see \citealt{zahn77-1} and \citealt{Mustill_2012} for details): 

\begin{equation}
    \label{main_eq}
    \left(\frac{\dot{a}}{a}\right)=-\left(\frac{\dot{M}_{\star}}{M_{\star}+M_{\mathrm{p}}}\right)-\left(\frac{\dot{a}}{a}\right)_{\mathrm{t}},
\end{equation}
where $M_{\mathrm{p}}$ is the mass of the substellar object, $M_{\star}$ is the total mass of the star, and  $a$ is the semi-major axis. The first term describes the orbital expansion due to stellar mass loss, which is assumed to be isotropic. Mass loss from the companion is not included, as it is expected to be negligible compared to the stellar mass-loss rate  \citep{Villaver_2009, schreiberetal19-1}. The term $(\dot{a}/a)_{\mathrm{t}}$ represents the change of the orbital distance caused by stellar tides and is given by
\begin{equation}
    \label{stellar_tides}
    \begin{aligned}
    \left(\frac{\dot{a}}{a}\right)_{\mathrm{t}}= & -\frac{1}{9 \tau_{\text{conv}}} \frac{M_{\star}^{\mathrm{env}}}{M_{\star}} \frac{M_{\mathrm{p}}}{M_{\star}}\left(1+\frac{M_{\mathrm{p}}}{M_{\star}}\right)\left(\frac{R_{\star}}{a}\right)^8 \\
    & \times\left[2 p_2+e^2\left(\frac{7}{8} p_1-10 p_2+\frac{441}{8} p_3\right)\right].
\end{aligned}
\end{equation}
Similarly, the eccentricity of the orbit is changing as
\begin{equation}
\label{eq:eccentricity}
\begin{aligned}
\left(\frac{\dot{e}}{e}\right)_{\mathrm{t}} 
= & -\frac{1}{36\,\tau_{\mathrm{conv}}} 
\frac{M_\star^{\mathrm{env}}}{M_\star} 
\frac{M_{\mathrm{p}}}{M_\star}
\left( 1 + \frac{M_{\mathrm{p}}}{M_\star} \right)
\left( \frac{R_\star}{a} \right)^8  \\
& \times \left[ \frac{5}{4}\,p_1 - 2\,p_2 + \frac{147}{4}\,p_3 \right].
\end{aligned}
\end{equation}
Here, $M_{\star}^{\mathrm{env}}$ is the envelope mass, $R_{\star}$ is the radius of the star, $e$ is the eccentricity of the orbit, $\tau_{\mathrm{conv}}$ is the eddy turnover timescale, and $p_{i}$ 
represents the frequency components of the tidal force according to \citet{Mustill_2012}, that is, 
\begin{equation}
    \label{tidal_comp}
    p_i \approx \frac{9}{2} \min \left[1,~\left(\frac{4 \pi^2 a^3}{i^2 G\left(M_{\star}+M_{\mathrm{p}}\right) \tau_{\mathrm{conv }}^2}\right)\right], ~ i=1,~2,~3.
\end{equation}
Given that the convective turnover time is uncertain,   
we considered two different expressions for $\tau_{\mathrm{conv}}$. 
The first one was presented in \citet[][hereafter R96]{rasioetal96-1} and is given by
\begin{equation}
    \label{Rasio_tconv}
    \tau_{\mathrm{conv}} = \left( \frac{M_{\star}^{\mathrm{env}} (R_\star - R_{\star}^{\mathrm{env}}) R_{\star}^{\mathrm{env}}}{3 L_\star} \right)^{1/3}
,\end{equation}
while an alternative prescription was proposed more recently by  \citet[][hereafter VL09]{Villaver_2009}, that is, 
\begin{equation}
    \label{Villaver_tconv}
    \tau_{\mathrm{conv}} = \left( \frac{M_{\star}^{\mathrm{env}} (R_\star - R_{\star}^{\mathrm{env}})^{2} }{3 L_\star} \right)^{1/3}.
\end{equation}
In both Eq.\,\ref{Rasio_tconv} and Eq.\,\ref{Villaver_tconv}, $R_{\star}^{\mathrm{env}}$ is the radius at the base of the convective envelope and $L_\star$ is the luminosity of the star.
The two prescriptions for the convective turnover time lead to different results, especially for extended convective envelopes. The turnover times from \citetalias{rasioetal96-1} are shorter, which implies stronger stellar tides for stars on the RGB and the AGB.

\begin{figure*}
    \centering
    \begin{minipage}{1.0\textwidth}
        \centering
        \includegraphics[width=\textwidth]{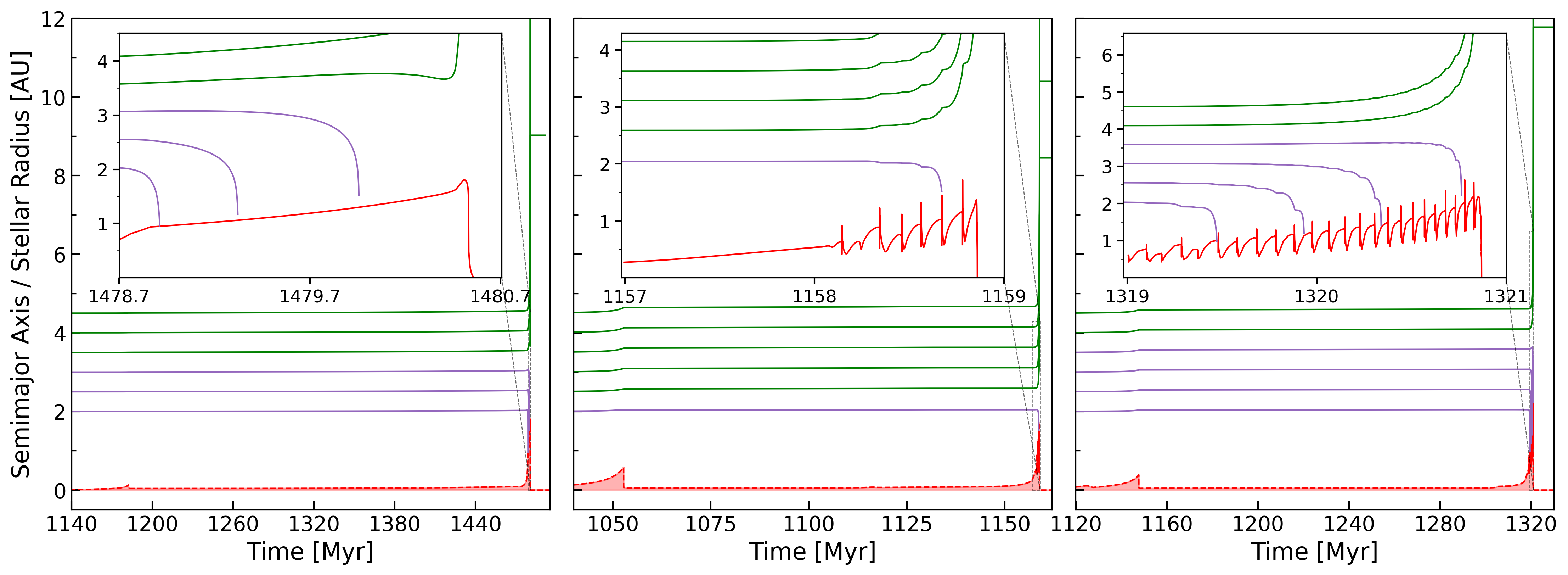}
    \end{minipage}
    \caption{
    Orbital evolution of a $1$\,\Mjup~gas-giant planet orbiting a $2$\,\Msun star with $Z=0.0187$, comparing calculations performed with SSE (left), the MESA default test suite (middle), for which the AGB mass-loss efficiency is set to $\eta=0.7$, and MESA assuming a more realistic mass-loss efficiency ($\eta= 0.02$; right).
    The orbits are assumed to be circular, and initial separations range from $2$ to $4.5$\,au, with a step size of $0.5$\,au. The red filled area corresponds to the stellar radius, while purple and green lines denote the orbital separation of the engulfed and surviving planets, respectively. The insets show zoomed-in views of the thermal pulse phase of the AGB and highlight its critical role in planetary engulfment. SSE does not account for the thermal pulses (left). For a large mass-loss efficiency (middle), most planets survive because the orbit expansion caused by stellar mass loss dominates. In the most realistic scenario assuming a small efficiency (right), the star evolves through more thermal pulses and reaches a larger radius, which causes most planets to be engulfed.}
    \label{fig:stellar_evolution}
\end{figure*}

\begin{figure}
    
     \resizebox{\hsize}{!}{\includegraphics{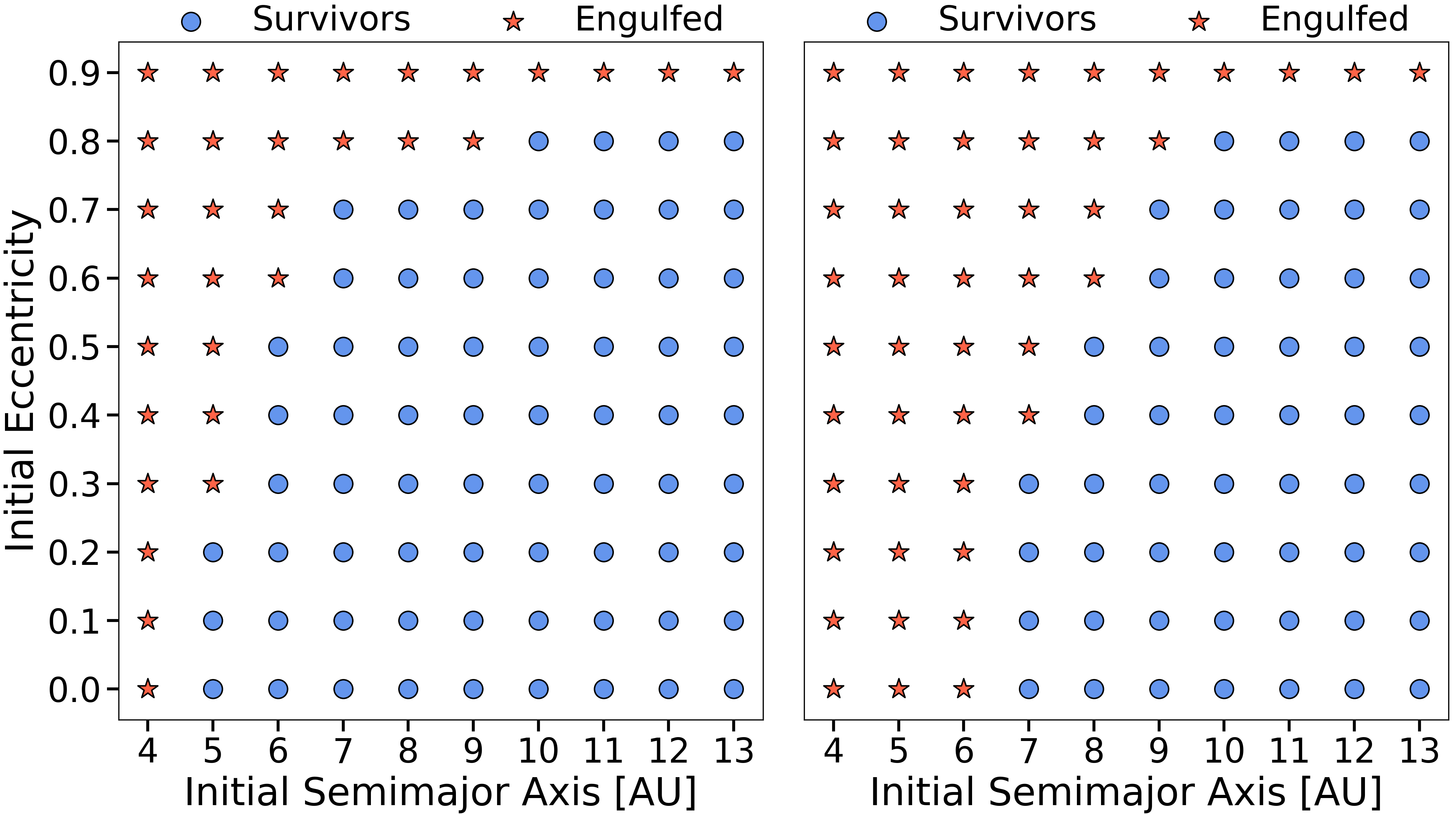}}
   
    \caption{Survival of a $10$\,\Mjup~gas-giant planet orbiting a $2$\,\Msun~star for different initial semimajor axes and eccentricities.  
    Red stars indicate engulfment, while blue circles stand for survival of the planet.  
    We assumed two different approximations for the convective turnover time, $\tau_{\mathrm{conv}}$. Following \citetalias[][]{rasioetal96-1} (right panel), many more planets are engulfed than when assuming the prescription suggested by \citetalias[][]{Villaver_2009} (left panel). 
    In both cases, the survival of substellar objects depends on the eccentricity of the orbit, with more eccentric orbits leading more frequently to engulfment. Due to stronger tides, this dependence is more pronounced assuming the approximation by \citetalias[][]{rasioetal96-1}. For initial semimajor axes $\geq8$\,au, the effects of tidal forces are negligible except for very large eccentricities. 
    }
    \label{fig:tides+eccentricity}
\end{figure}

\subsection{Impact of stellar evolution codes, stellar tides, mass-loss efficiency, and eccentricity}

Before carrying out population synthesis for substellar companions to WDs, we tested the impact of stellar evolution and stellar tides on the survival of these companions. As we show here, the modeling of mass loss on the AGB phase and tides can be crucial. 

As in most stellar evolution codes, we modeled mass loss on the first giant branch with the prescription by \citet{Reimers1975} with an efficiency of $\eta=0.5$. Mass loss on the AGB is more uncertain. The two preferred prescriptions are those proposed by \citet{Bloecker95-1} and \citet{vassiliadis+wood93-1}. Originally, \citet{Bloecker95-1} proposed an efficiency on the order of one, but very different scaling factors have been suggested since then, including dependencies of the efficiency on mass, metallicity, or evolutionary state \citep{pignatarietal16-1,huscheretal25-1}. We reflect this general uncertainty by using the prescription from \citet{Bloecker95-1} with two different (but constant) values for the mass-loss efficiencies ($\eta=0.02$ and $\eta=0.7$). The lower value is probably more realistic \citep{ventura-12,Cinquegranaetal-22}. We therefore employed the former value for our baseline simulations.

To illustrate the impact of different mass-loss prescriptions and stellar evolution models, we simulated the orbital evolution of a $1$\,\Mjup\,giant planet orbiting a $2$\,\Msun~star with nearly solar metallicity ($Z = 0.0187$), considering a range of initial orbital separations. For simplicity, in this exercise all orbits were assumed to be circular. Stellar evolution was modeled with the simple single star evolution 
(SSE) code \citep{hurleyetal00-1} that does not include thermal pulses on the AGB and MESA assuming $\eta=0.02$ and $\eta=0.7$. 

Figure~\ref{fig:stellar_evolution} presents the evolution of the giant planet's orbital separation calculated  
under these three scenarios, assuming the turnover timescale from \citetalias{Villaver_2009}. Inspecting the two panels with MESA simulations, we note that a higher value for the mass-loss efficiency (middle panel) leads to fewer thermal pulses during the AGB, which in turn limits the maximum stellar radius. As a result, companions can survive at smaller initial separations. Conversely, lower mass-loss efficiencies lead to larger maximum radii, requiring significantly wider orbits to avoid engulfment.
The results obtained using the SSE code predict a similar maximum radius to that obtained under high-efficiency mass loss in MESA. Most importantly, SSE does not describe the radius evolution of thermally pulsing AGB stars, which is key for determining the survival of companions. 

A second uncertainty that needs to be considered concerns the prescriptions used for stellar tides. 
In Fig.\,\ref{fig:tides+eccentricity}, we highlight how different prescriptions for the convective turnover timescale, which impact the strength of stellar tides, affect the survival of substellar companions for different orbital eccentricities. Stellar tides are stronger assuming the approximation of the turnover time proposed by \citetalias{rasioetal96-1}, which implies that substellar companions need to be located at larger initial separations to survive. 
In general, the larger the eccentricity of the orbit, the larger the semimajor axis needs to be for survival. This is because, for highly eccentric orbits, stellar tides can start to dominate during periastron and may ultimately cause engulfment.

The examples shown in Figs.\,\ref{fig:stellar_evolution} and \ref{fig:tides+eccentricity} illustrate that the survival of substellar objects up to the WD phase depends on the assumptions made for stellar evolution on the AGB as well as on the assumed approximation for the convective turnover timescale. We took these findings into account when designing our population-synthesis approach. 

\section{Population synthesis} 

To theoretically predict the population of substellar objects orbiting WDs, we first generated a sample of potential WD progenitor stars taking into account the initial mass function and the age-metallicity relation.  
We further assumed a companion occurrence-rate distribution (depending on mass and metallicity of the host star) consistent with radial-velocity surveys of $0.8-4$\,\Msun~giant stars \citep{wolthoff2022}. We then simulated the future evolution of these planetary systems to estimate the present-day population of substellar companions to WDs. In the following sections, we detail our population-synthesis methodology.

\subsection{Initial distributions of potential WD progenitors}

The population of potential WD progenitors was generated using the initial mass function from \citet{kroupa2001} for the mass range from $0.8$ to $4$\,\Msun. We excluded larger masses because the substellar companions' occurrence rate is unconstrained. This also means that our predicted sample excludes WD masses exceeding $\sim0.90$\,\Msun. We generated $243\,617$ stars. 

We then assigned each star an age assuming a constant star formation rate using $10$\,Gyr as the age of the thin disk of the Milky Way. 
The metallicity of each star was incorporated taking into account 
an age-metallicity relation by interpolating the parameters defined in \citet[][their Table 1]{carrillo2023}. 
This age-metallicity relation was derived from evolutionary simulations of the Milky Way. Therefore our population synthesis does apply to global populations of WDs and not necessarily to local observed populations where the age-metallicity relation might have evolved differently \citep{Fernandez-Alvar-2025}.

\subsection{Probability distributions for substellar objects around potential WD progenitors}

The progenitor stars of the current WD population were intermediate-mass stars (i.e., $\sim0.8-8$\,\Msun). Detecting substellar companions around main-sequence stars with masses exceeding $\sim1.5$\,\Msun~is challenging as the absence of a large number of absorption lines makes radial-velocity studies extremely difficult, and the larger contrast complicates the direct detection through imaging and transits.

Reliable constraints on the occurrence rate of giant planets and brown dwarfs at large orbital separations around stars that can evolve into a WD within the age of our Galaxy have been derived from radial-velocity studies of giant stars \citep{reffertetal15-1,wolthoff2022}. 
We used their probability distributions which cover the companion mass range from $0.8$ to $30$\,\Mjup~to generate a representative population of substellar companions around the potential WD progenitors, that is, 
we assumed that the companion occurrence rate ($f$) depends on stellar mass and metallicity as
\begin{equation}
        f\left(M_*,[\mathrm{Fe/H}]\right)=C\cdot\exp\left(-\frac{1}{2}\left[ \frac{M_*-\mu}{\sigma}\right]^2\right)\cdot10^{\beta\cdot [\mathrm{Fe/H}]}
    \label{eq:occ}
,\end{equation}
with 
\begin{align*}
        C &= \left(0.128\pm 0.027\right)\\
        \mu &=\left(1.68\pm 0.59\right)\,M_\sun\\
        \sigma &=\left(0.70\pm 1.02\right)\,M_\sun\\
        \beta &=\left(1.38\pm 0.36\right). 
\end{align*} 
The orbital period distribution of the companions is given by \citep{wolthoff2022}
\begin{align}
        f_\mathrm{occ}\left(P\right)=\frac{A}{\sqrt{2\pi} \sigma \cdot P/\mathrm{days}}\cdot \exp\left(-\frac{\left(\ln(P/\mathrm{days})-\mu\right)^2}{2\sigma^2}\right), 
\end{align}
with 
\begin{align*}
        A&=\left(9988\pm 4166\right)\\
        \sigma &=\left(0.89\pm 0.19\right)\\
        \mu &=\left(7.46\pm 0.45\right).
\end{align*}
We adopted the distribution recently obtained by \citet{Kane24} for the eccentricities of the orbits of the companions. 
Excluding planets with very short periods ($P<10$ days), these authors showed that the eccentricity distribution of giant planets has a mean value of $0.29$, a median of $0.23$, and a $1\sigma$ RMS scatter of $0.23$. The distribution is concentrated toward low eccentricities, but it extends up to $e\approx0.9$. These high eccentricities are thought to result from planet-planet scattering and/or the Kozai-Lidov effect \citep{Alqasimetal25}.

Finally, we assumed a planet mass function derived from observations of companions to FGK main-sequence stars, which indicate that the mass distribution for companions in the range of $0.3M_{\text{J}}\leq M_{\text{p}}\leq 10 M_{\text{J}}$ can be reasonably approximated by a power law \citep[$\propto M_{\mathrm{p}}^{-1.3}$,\,][]{cummingetal08-1,mayoretal11-1,adamsetal21-1}, and we extrapolated this distribution up to $30$\,\Mjup. 
While the number of companions detected by \citet{wolthoff2022} was too small to derive a planet mass function, our approach seems to be roughly in agreement with their finding that the number of companions decreases with mass (see their Fig. 7). We furthermore note that by extrapolating the planet mass function up to the brown dwarf regime, we probably overestimated their occurrence. This is because the mass range between $13$ and $30$~\Mjup~at separations of a few astronomical units encompasses the brown-dwarf desert, where such companions have been shown to be very rare \citep[see, e.g.,][]{Grether2006,Vanzandt2025}. However, we estimate that this discrepancy affects our results by $\sim10\%$ at most, which is significantly smaller than the uncertainties associated with the other adopted distributions.

Using the distributions described above, we assigned a representative number of substellar companions to the population of potential WD progenitors ($0.8-4$\,\Msun) using a Monte Carlo sampling approach.
The resulting fraction of these stars hosting a companion was $5.9\pm3.4$ \%. 
To predict a present-day population of substellar objects around WDs, each star in the population we generated was evolved up to today, taking into account the impact of stellar evolution on the substellar companion orbit, if present.

\subsection{Generating a representative WD sample}

Equipped with a sample of stars that are potential progenitors of present-day WDs (according to their mass), knowing their metallicity and age, and having assigned a representative population of substellar companions, 
we simulated the evolution of these systems in order to predict a
representative population of present-day WDs including substellar companions. 

Given that the sample of stars that we generated was far too large to perform each simulation with MESA even in a super-cluster, we decided to evolve all stars from their birth time to today using SSE. We found that of the initial sample of 243\,617 stars ($16\,086$ with substellar companions), $113\,985$ are predicted to be present-day WDs ($5611$ of them initially hosting a companion). 
While evolving the host stars with the fast stellar evolution code SSE should not introduce significant errors --the ages and masses of the resulting WDs are similar to those predicted with MESA-- 
the survival of the substellar companions depends on the details of late AGB evolution, especially the thermal pulses and the mass loss.  The details of both processes are not well described by SSE (see Fig.\,\ref{fig:stellar_evolution} for details), and we therefore used more detailed stellar evolution tracks to estimate the survival of the companions.  

We randomly selected $1000$ stars that have become WDs (according to our SSE simulations) and initially hosted a substellar companion, and we simulated the evolution of the host star with MESA. The selection of $1000$\,stars corresponds to a fraction of $17.82$\,\% of all stars that initially hosted a companion and are currently WDs. To establish our final representative sample of present-day WDs, we therefore randomly selected the same fraction of objects from the sample of stars that have become WDs but did not initially host a companion. This led to predicting $19\,315$ present day WDs. 
As the final step, we evaluated the survival of the $1000$ substellar companions using the code FATES and the stellar evolution tracks calculated with MESA.

\begin{figure*}
\sidecaption
    \includegraphics[width=12cm]{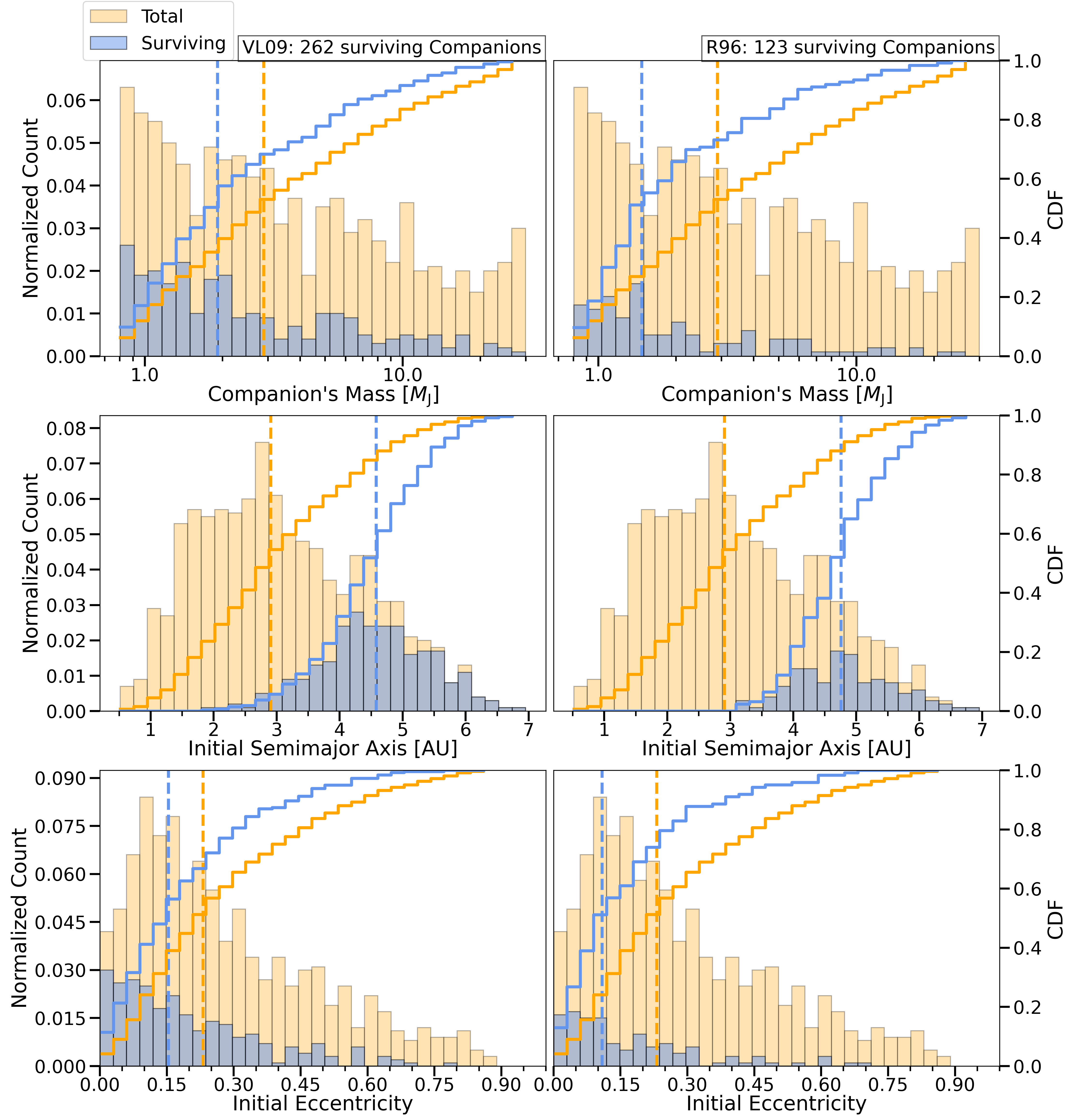}
        \caption{Survival of substellar companions as function of companion mass (top), initial semimajor axis (middle), and initial eccentricity (bottom). 
        The left panels represent results when considering the tidal force model by \citetalias{Villaver_2009}, while the right panels show results when using the tidal force model by \citetalias{rasioetal96-1}.  All distributions are normalized by
         the total number of stars (1000).
        The number of surviving companions is indicated in the figure for each model (blue histogram). The vertical lines correspond to the median value of each sample. The solid lines correspond to the cumulative distributions. 
        In general, the initial distributions of surviving companions (orange) are shifted toward lower masses, larger semimajor axes, and smaller eccentricities compared to the initial values of all companions (blue).  
        These effects are slightly more pronounced for stronger tides (right panels). 
        }
        \label{fig:Initial planets distributions}
\end{figure*}

\section{Results}

Our final sample consists of $20\,315$ WDs whose parameters were calculated with SSE, and $1000$ of these WDs initially hosted a companion whose final orbital parameters were calculated using the \textit{FATES} code from \citet{roncoetal20-1} with stellar evolution tracks from MESA. 
We first examine the outcomes for the $1000$\,stars that initially hosted a companion.

\subsection{Survival and initial companion characteristics}
\label{sec:survival}

We simulated the orbital evolution of the $1000$ substellar companions assuming the approximations of the convective turnover time from \citetalias{rasioetal96-1} and \citetalias{Villaver_2009}. In the latter case, the resulting stellar tides are weaker, and more companions are therefore expected to survive. 

The top panels of Fig.\,\ref{fig:Initial planets distributions} show that the rate of surviving companions indeed depends on the assumed prescription for the convective turnover time. 
While more than a quarter of all substellar companions survived ($26.2$\,\%) assuming the approximation by \citetalias{Villaver_2009}, only $12.3$\,\% survived if
the prescription for the turnover timescale from \citetalias{rasioetal96-1} was assumed.
As more massive companions generate stronger tides, the median of the surviving companion mass distribution is shifted somewhat to smaller companion masses. Of the population of surviving companions, $93.5$ \% would be classified as giant planets; i.e., they would have masses below $\sim 13$\Mjup~\citep[][]{burrowsetal97} when weak tides are assumed, compared to $95.1$\,\% under strong tides. 

Similarly, the initial semimajor axis has to be somewhat larger on average for companions to survive in the case of stronger tides (middle panels of Fig.\,\ref{fig:Initial planets distributions}). Assuming weak tides, the initial semimajor axis of some surviving companions is $\sim2.5$\,au (left), while the closest substellar companions that survive in the case of strong tides are initially located slightly farther away from the star ($3$\,au).      
  
Finally, companions with a larger initial eccentricity are more likely to be engulfed. 
This is because for a given orbital separation, the larger the eccentricity of the companion, the closer to the host star the companion is during periastron, which can lead to significantly increased tides. 
Therefore, surviving companions have smaller eccentricities on average. 
Again, this effect is more pronounced when assuming stronger tides 
(bottom panels of Fig.\,\ref{fig:Initial planets distributions}).

\subsection{Characterizing WDs and their companions}

\begin{figure*}
  \begin{center}
    \includegraphics[width=\linewidth]{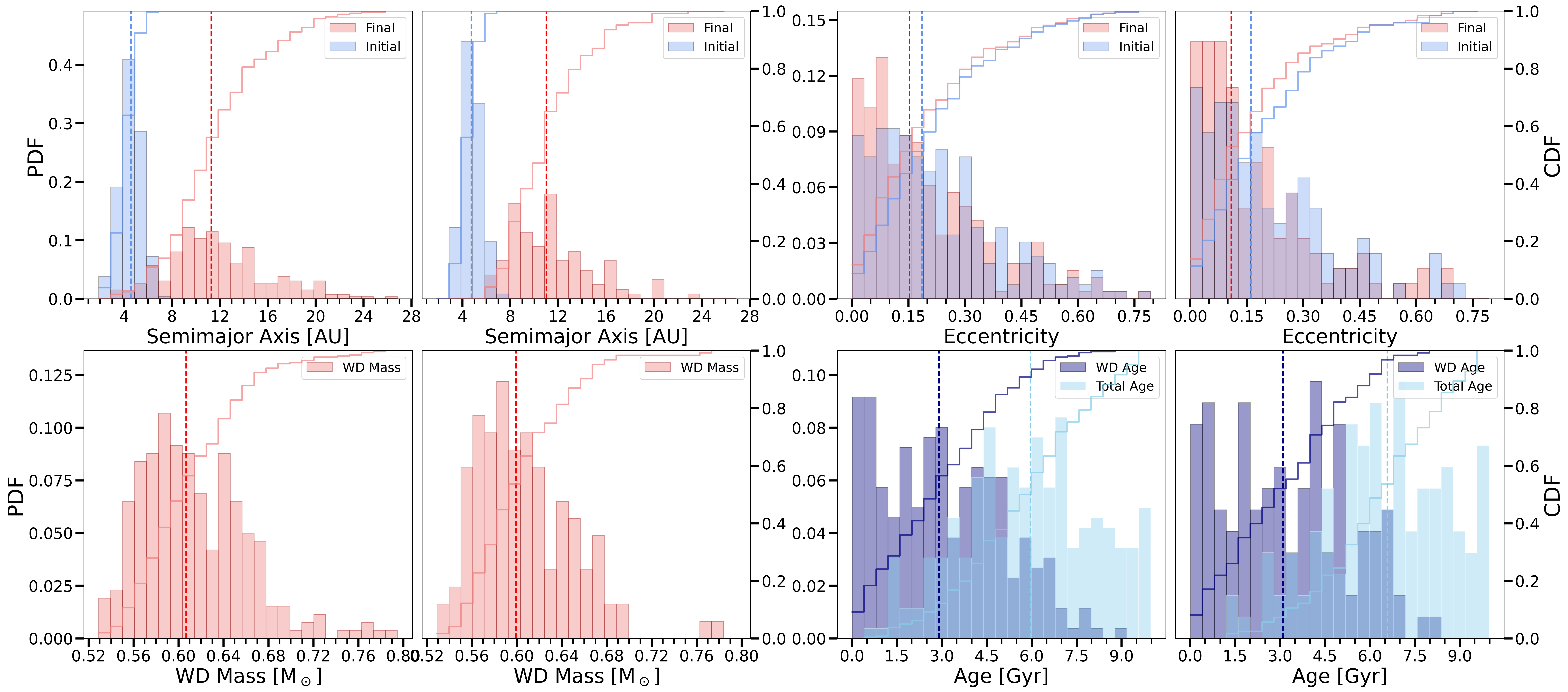}
  \end{center}
  \caption{Final and initial properties of systems that host a susbtellar companion. Each row shows a set of histograms comparing two WD samples: one modeled using \citetalias[][]{Villaver_2009} (left) and one based on the \citetalias[][]{rasioetal96-1} (right) approximation for stellar tides. All histograms are normalized to the total number of surviving companions for each simulation. The final semimajor axes are significantly larger than the initial ones for both tidal prescriptions (top left). For weak tides, the eccentricity distribution does not significantly change, while the distribution moves slightly toward smaller eccentricities for strong tides (top right). The bottom panels show the WD mass and age distributions for both approximations of tidal forces which appear to be rather similar.}
  \label{fig:final-and-WD}
\end{figure*}

Having inspected the relation between the initial distributions of companions that survive or are engulfed for both prescriptions of the convective turnover time, we now take a detailed look at the predicted final semimajor axis and eccentricity of the surviving companions and the properties of their host stars. Again, we present the results for weaker and stronger tides resulting from the different prescriptions for the convective turnover time.  

In the top panels of Fig.\,\ref{fig:final-and-WD}, we show the distributions of the final semimajor axis and eccentricity of the surviving companions and compare them with the initial ones. 
Independently of the strength of the stellar tides, the semimajor axis of the surviving companions increases significantly, on average, due to the evolution of the host star into a WD. In addition, the distributions become much broader (top left panels of Fig.\,\ref{fig:final-and-WD}). This is easy to understand. For most surviving companions, mass loss dominates, and, consequently, the orbital separation generally increases. 
We note, however, that changes in semimajor axis depend not only on their initial value and the eccentricity, but also on the mass of the central star and the companion. In addition, for a few companions the semimajor axis increases only slightly, or even not at all, as the dominance of mass loss over stellar tides is much less pronounced. All these effects combined 
explain why the resulting final distributions are much broader than the initial one. 

The strength of stellar tides also plays a role when comparing the initial and final eccentricities of the surviving companions.  
Eccentricities become significantly smaller due to stellar tides in the case of stronger tides. This effect mostly results from the fact that companions with initially intermediate eccentricities ($\sim0.15-0.30$) are reduced due to stellar tides. Companions with an initially very large eccentricity are either engulfed during the giant phases or have initial semimajor axes large enough that 
even during periastron passage stellar tides remain relatively inefficient (see Fig.\,\ref{fig:final-and-WD}, upper right panels). We therefore predict that a non-negligible fraction of the surviving companion population may retain eccentric orbits (especially under weak tides), which could contribute to the perturbation of smaller bodies, either delivering them into the WD or leading to their ejection \citep{Frewen2014}.

The masses of the predicted WD population with surviving companions are quite similar for stronger and weaker stellar tides. In both cases the distribution peaks at typical single-WD masses of $\sim0.6$\,\Msun. This is largely because low-mass progenitor stars are more frequent and because the substellar companion occurrence rate peaks around $1.7$\,\Msun. In both cases, very few massive WDs $(\sim0.8$\,\Msun) are predicted to host companions. 
These WDs must have formed from low-metallicity stars of $\sim3-3.5$\,\Msun\,or even more massive ($3.5-4$\,\Msun) high-metallicity stars \citep[see][Fig. 3]{baueretal25-1}. We note that for this mass range the occurrence rate of companions is highly uncertain (see \citealt{wolthoff2022} for details).

In the bottom right panel of Fig.\,\ref{fig:final-and-WD}, we show the age distributions for predicted WDs with companions. In general, companions are more likely around young WDs because the age-metallicity relation implies a larger companion occurrence rate for younger stars. The total age of the companions hosts, however, peaks at around $5$\,Gyr, because younger stars in our potential progenitor sample had often not yet evolved into WDs.   

\subsection{Fractions of WDs with companions}

\begin{figure*}
  \begin{center}
    \includegraphics[width=0.82\linewidth]{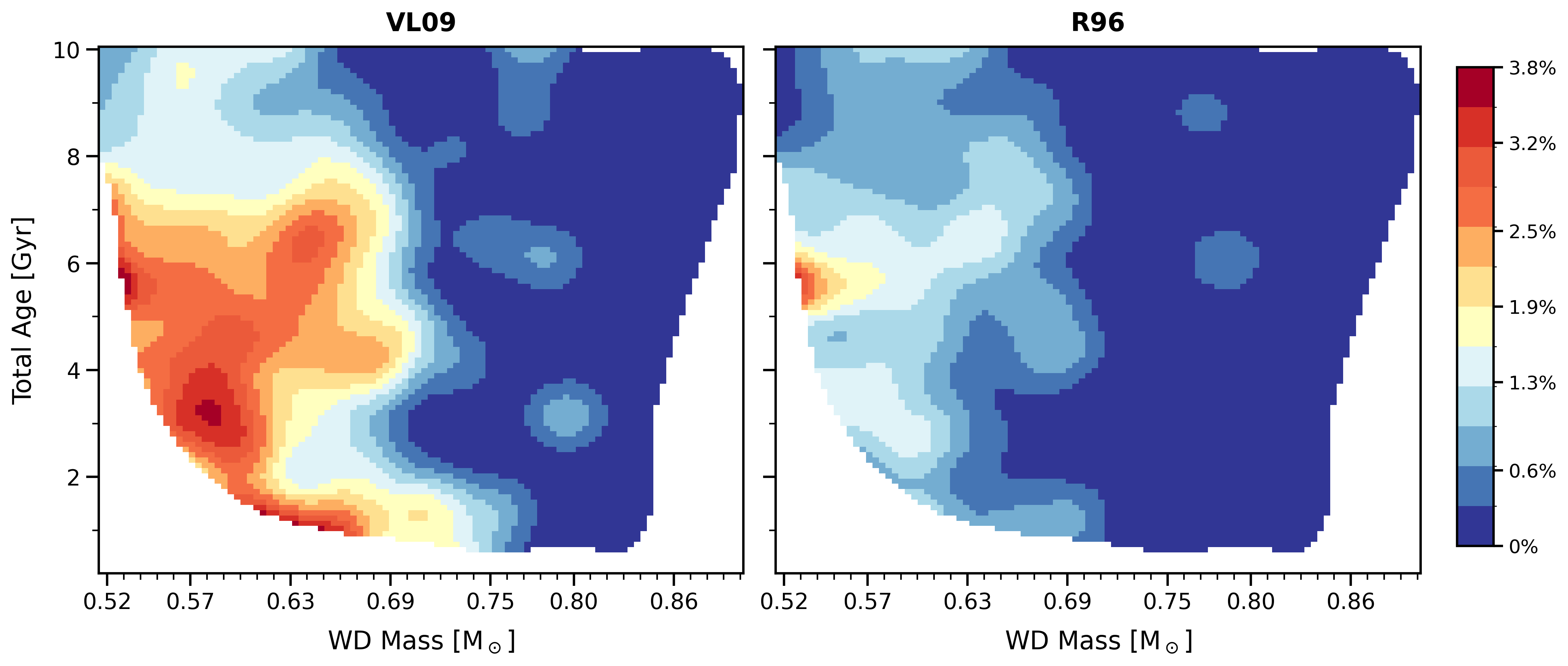}
  \end{center}
  \caption{Fractions of WDs hosting substellar companions as function of total system age and WD Mass. The total fraction of WDs with substellar companions is $1.3$ \% \citepalias[assuming weak tides][top left]{Villaver_2009} and $0.6$ \% for stronger tides according to \citetalias{rasioetal96-1}. In both cases, companions are more likely around lower mass WDs and relatively young systems. In particular, for higher mass stars, stronger tides lead the radius expansion to dominate over orbital expansion, and therefore only planets around lower mass stars survive. The extension of the region with high probabilities is therefore very sensitive to the strengths of the tides. For example, further increasing the tides according to \citetalias{rasioetal96-1} by a factor of $8/3$ would eliminate the remaining small high probability island in the right panel. }
  \label{fig:four_panel_KDE}
\end{figure*}

The perhaps most important question we would like to answer through our population synthesis is how many WDs are expected to host substellar companions. In order to answer this question, we evolved a representative number of WD progenitors, independently of whether they host a companion, from their assigned birth time to the present day (using SSE). The survival and the final orbital parameters were determined using stellar evolution tracks from MESA. The resulting population of WDs allowed us to determine the relative number of WDs hosting substellar companions. 

We find that in total $1.3$\,\% of the predicted WD population hosts a companion if weak tidal forces according to the turnover time approximation by \citetalias{Villaver_2009} are assumed. This number decreases to $0.6$\,\% if stronger tides resulting from the turnover times by \citetalias{rasioetal96-1} are assumed.   
Of the population of surviving companions, 93.5\,\% would be classified
as giant planets when weak tides are assumed,
compared to 95.1\,\% under strong tides. Therefore, the resulting global occurrence of giant planets
around WDs would be approximately 1.2\,\% for the weak-tide case and 0.6\,\% for case
of the strong tides. In other words, the vast majority of the predicted substellar companions are gas giant planets. 

Given that the strength of stellar tides is uncertain and both prescriptions tested here could be poor descriptions of reality, as an additional test we simulated the evolution of the $1000$ substellar companions without tidal forces to explore the maximum survival rate when the orbital evolution is governed solely by stellar mass loss. We find that the survival rate increases significantly (see third row of Table \ref{tab:summary_results}), yielding a global occurrence of $3.2$\,\%. This difference clearly shows how important stellar tides are in shaping some properties of the surviving companions population.

In addition to estimating the total fraction of WDs hosting substellar companions, we can evaluate how this fraction depends on the characteristics of the WD and its progenitor. 
In Fig.\,\ref{fig:four_panel_KDE}, we show the fraction of WDs with companions as a function of the WD mass and the total age of the system. For both assumed tidal forces, the predicted companion occurrence rate is largest for young ($\sim1-6$\,Gyr) and relatively low-mass ($\sim0.53-0.66$\,\Msun) WDs. This result can be understood by the dependence of the companion occurrence rate on stellar mass (which peaks around $\sim1.7$\,\Msun) and metallicity, which (according to our assumed metallicity-age relation) increases with age.  

While the general tendencies seem to be independent of the efficiency of stellar tides, we note that the maximum occurrence rate slightly shifted toward older stars and smaller masses if strong tides were assumed (middle panel of Fig.\,\ref{fig:four_panel_KDE}). 
This small shift is related to the fact that for strong tides and large stellar masses only companions at very large semimajor axes survive (as also illustrated in Fig.\,\ref{fig:final-and-WD}, lower left panels).

Finally, in order to estimate the statistical uncertainty of the predicted fractions of WDs with companions, we generated $3000$ populations of WDs ($\sim100\,000$ WDs each) with potential companions, ignoring the possible effects of stellar evolution but varying the parameters in the occurrence-rate function (Eq.\,\ref{eq:occ}) according to their uncertainties. 
We find that without taking into account the impact of stellar evolution (i.e., assuming that all companions survive), we can predict an occurrence rate of $4.4\pm2.4$ \% ($1\sigma$ uncertainty). This value represents a strict upper limit. 
The relatively large uncertainty arises mainly from higher stellar masses ($M_{\text{MS}}>2.5M_\odot$), where the companion occurrence is highly uncertain (see Eq. \ref{eq:occ}). 
This implies that the statistical error of the obtained fractions should generally be roughly $50$\,\%. 

In summary, our simulations predict that the substellar companion occurrence rate around WDs, even if only weak tides are assumed, does not exceed $\sim1.3\pm0.7$ \% and that the expected fraction is largest for low-mass WDs with total ages below $\sim6$\, Gyr. However, this result depends on several assumptions that we made in our population-synthesis simulations. In what follows, we investigate to what degree our results depend on these assumptions and what our results might imply for the observability of substellar companions around WDs.

\begin{table*}[htbp]
\centering
\renewcommand{\arraystretch}{1.5} 
\caption{Predicted companion population around WDs for different model assumptions.} 
\label{tab:summary_results}
\begin{tabular}{lccccccc}
\hline\hline
1000 Planets
Simulation & $a_f$ & $e_f$ & $M_\mathrm{WD}$ & $t_\mathrm{WD}$ & $t_\mathrm{tot}$ & $f_\mathrm{surv}$ & $f_\mathrm{WD}$ \\
           & (AU) &  & (M$_\odot$) & (Gyr) & (Gyr) &  & \\
\hline
Villaver (Baseline)& $11.3_{-3.2}^{+4.8}$ & $0.15_{-0.11}^{+0.20}$ & $0.61_{-0.04}^{+0.06}$ & $2.91_{-2.19}^{+2.22}$ & $5.96_{-2.41}^{+2.41}$ & 262/1000 & 1.3\% \\
Rasio (Baseline)& $11.0_{-2.7}^{+3.8}$ & $0.11_{-0.07}^{+0.19}$ & $0.60_{-0.03}^{+0.05}$ & $3.10_{-2.40}^{+2.51}$ & $6.58_{-2.27}^{+2.02}$ & 123/1000 & 0.6\% \\
No Tidal Forces & $9.6_{-3.4}^{+4.5}$ & $0.18_{-0.11}^{+0.18}$ & $0.61_{-0.04}^{+0.05}$ & $2.67_{-1.97}^{+2.46}$ & $5.48_{-2.26}^{+2.51}$ & 652/1000 & 3.2\% \\
Villaver ($P_{\text{break}}=1500$ days) & $15.4_{-5.5}^{+6.9}$ & $0.19_{-0.12}^{+0.25}$ & $0.61_{-0.04}^{+0.05}$ & $2.66_{-1.96}^{+2.45}$ & $5.53_{-2.37}^{+2.48}$ & 573/1000 & 2.8\% \\
Rasio ($P_{\text{break}}=1500$ days) & $17.0_{-6.4}^{+5.9}$ & $0.17_{-0.12}^{+0.24}$ & $0.61_{-0.04}^{+0.05}$ & $2.77_{-2.07}^{+2.35}$ & $5.75_{-2.20}^{+2.43}$ & 386/1000 & 1.9\% \\
Villaver ($e=0$) & $11.3_{-3.4}^{+4.2}$ & $0.00$ & $0.61_{-0.04}^{+0.05}$ & $2.80_{-2.07}^{+2.45}$ & $5.83_{-2.45}^{+2.39}$ & 343/1000 & 1.7\% \\
Rasio ($e=0$) & $11.4_{-2.8}^{+3.9}$ & $0.00$ & $0.60_{-0.03}^{+0.05}$ & $2.92_{-2.29}^{+2.21}$ & $6.17_{-2.39}^{+2.22}$ & 170/1000 & 0.8\% \\
Villaver  ($\eta_B = 0.7$) & $11.6_{-4.4}^{+5.8}$ & $0.19_{-0.12}^{+0.23}$ & $0.62_{-0.04}^{+0.05}$ & $2.70_{-1.98}^{+2.44}$ & $5.36_{-2.33}^{+2.60}$ & 579/1000 & 2.9\% \\
Rasio  ($\eta_B = 0.7$) & $12.9_{-4.3}^{+5.3}$ & $0.18_{-0.12}^{+0.26}$ & $0.62_{-0.04}^{+0.05}$ & $2.82_{-2.09}^{+2.32}$ & $5.34_{-2.20}^{+2.62}$ & 405/1000 & 2.0\% \\
\hline
\end{tabular}

\tablefoot{%
We tested the impact of the assumed mass-loss efficiency for stars on the AGB, the assumed strength of tidal forces, the adopted eccentricity and planet-mass distribution, as well as the age-metallicity relation assumed for the WD progenitor stars.
Each entry shows the median value with the 16th and 84th percentiles. 
Columns correspond to final semimajor axis, final eccentricity, WD mass, WD age, total system age, survival rate, and predicted fraction of WDs hosting substellar companions, normalized to the total sample of $20\,315$ WDs.
}
\end{table*}

\section{Discussion}

In our base models presented in the previous section we found an upper limit of $1.3\pm0.7$\,\% for WDs hosting substellar companions. The exact predicted fraction depends on the adopted prescriptions for tidal forces. The surviving companions are predicted to be located around intermediate-mass WDs with final semimajor axes covering a range of $3-24$\,au with a median of $11$\,au. 
These findings are based on the assumed initial companion distributions and the age-metallicity relation. In what follows, we address the reliability of our assumptions before briefly discussing the potential survival of engulfed companions and the observability of the predicted population of substellar objects around WDs. 

\subsection{Age-metallicity relation and star formation history}

\begin{figure*}
\sidecaption
  \includegraphics[width=12cm]{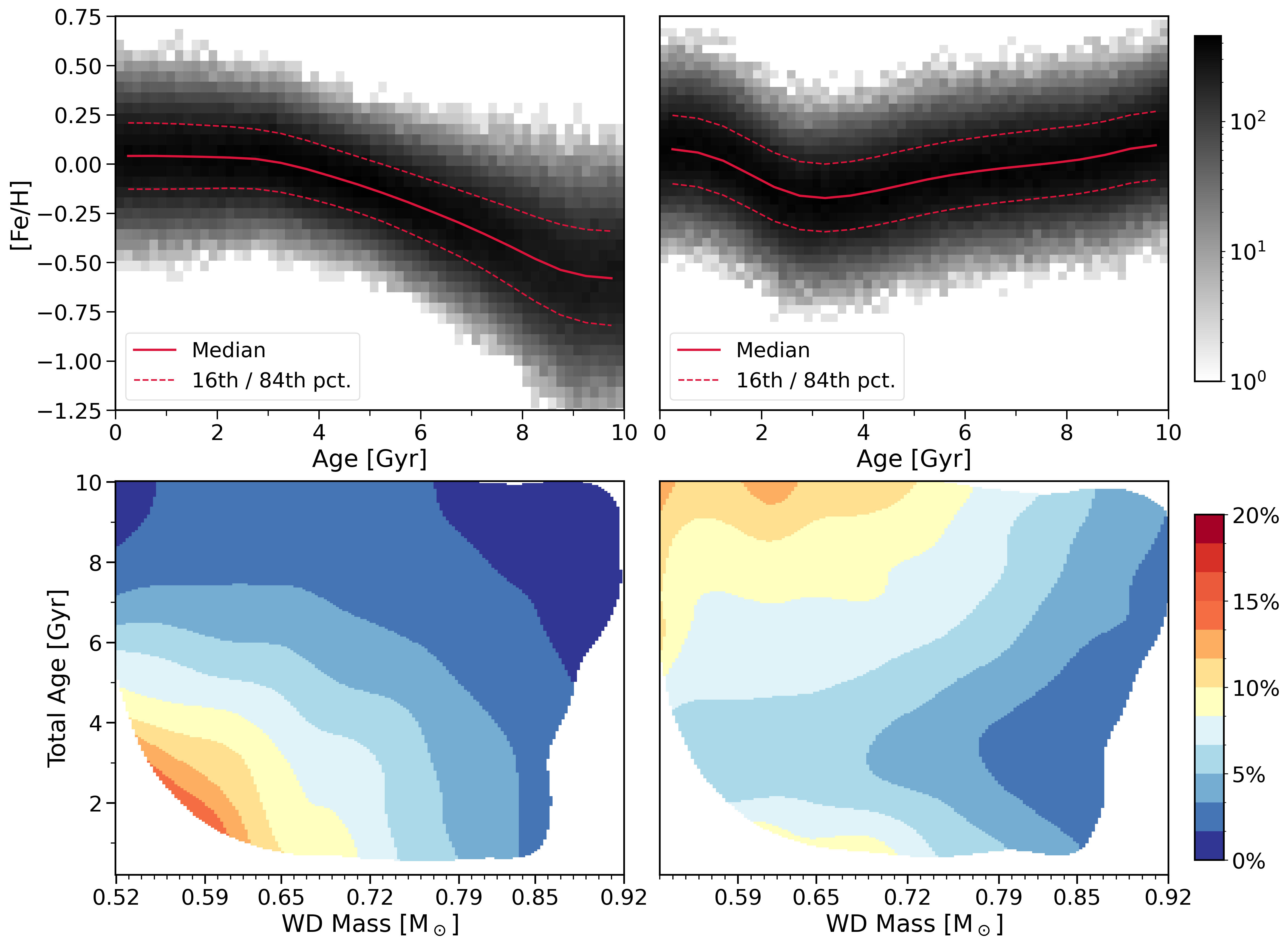}
    \caption{Age–iron abundance [Fe/H] relation for 243\,617 stars generated by interpolating distributions obtained by \citet{carrillo2023} (left panel) and for the estimate based on Fig.\,4 from \citet{Fernandez-Alvar-2025} (right panel; assuming the same dispersion as our baseline age-metallicity relation). The age-metallicity relation plays an important role in our population synthesis as the occurrence of substellar companions depends strongly on stellar metallicity. The strict upper limit (assuming all companions survive stellar evolution) for the occurrence rate of companions to white dwarfs increases by a factor of $\sim2$ (from $4.4\pm2.4$ to $8.2\pm 4.4 $\,\%), and old WDs are much more likely to host a companion when we assumed the age-metallicity relation from \citet{Fernandez-Alvar-2025}. }
    \label{fig:age-met}
\end{figure*}

The substellar-companion occurrence rate depends significantly on the metallicity of the host star. We adopted an age-metallicity relation derived from \citet{carrillo2023}, who used simulations to predict a global age-metallicity relation.  
For most age ranges (from $10$ up to $4$ Gyr ago), they find a relatively smooth and well-known continuous trend: the older the star, the lower its metallicity. 
For stars younger than 4\,Gyr the relation is essentially flat, clustering roughly around solar metallicity (with a relatively large scatter). This is consistent with classical approaches considering the chemical enrichment of the interstellar medium. 

Alternative studies, however, have revealed that the Milky Way might have a more complex history. 
\citet{Ruiz-Lara2020} applied Gaia DR2 color-magnitude diagram (CMD) fitting for stars within $2$\,kpc of the Sun and found that the star formation rate has been slowly decreasing and perturbations to our galaxy from the Sagittarius dwarf galaxy triggered enhanced star formation rates, especially at three ages: $5.7$, $1.9$, and $1.0$\,Gyr. 
More recently, \citet{Gallartetal2024} and \citet{Fernandez-Alvar-2025} used improved CMD fitting to derive the star formation history and age-metallicity relation for different local samples. 
According to these works, the thin disk's metallicity in the defined local regions was super-solar at the beginning ($\sim8-10$\,Gyr ago), decreased 
until an age of $\sim3$\,Gyr, and then increased again to super-solar values \citep[][their Fig. 4]{Fernandez-Alvar-2025}. 

In principle, these recent findings could directly affect the predicted fraction of substellar companions around WDs in the local volume.   
In order to estimate the potential impact of different star formation histories and age-metallicity relations, we performed two separate tests incorporating the age-metallicity relation (see right panel of Fig. \ref{fig:age-met} and star formation history from \citet{Fernandez-Alvar-2025}. In both cases, we predicted the population of WDs with substellar companions ignoring the impact of stellar evolution (because of the tremendous computational costs of running a meaningful number of MESA simulations). The obtained results therefore have to be taken as strict upper limits and should be compared with the corresponding upper limits predicted by our default model (constant star formation rate and standard age-metallicity relation). 

While changing the star formation history has very little impact on the predicted population of substellar companions around WDs (strict upper limit slightly increases from $4.4\pm2.4$ to $4.7\pm2.6$\,\%), given the strong dependence of the companion occurrence rate on metallicity, a change in the age-metallicity relation affects the prediction significantly. 
In the top panels of Fig.\,\ref{fig:age-met}, we compare the more local age-metallicity relation with that used in our population synthesis. Given that observational surveys will obviously target WDs close to the Sun, in this context the more local relation might provide more realistic predictions (right panel). 
Using the age-metallicity relation from \citet{Fernandez-Alvar-2025} increases the number of predicted companions by almost a factor of two (the strict upper limit --obtained by ignoring the fact that companions might be engulfed-- increases from $4.4\pm2.4$ to $8.2\pm4.4 $\,\%). 
The properties of WDs hosting companions would also be affected, as considering a super-solar metallicity at very old ages results in the prediction of many more old WDs hosting substellar companions
(see Fig.\,\ref{fig:age-met}, bottom panels). 
In summary, due to variations in the age-metallicity relation in local volumes or the Milky Way, the fraction of WDs hosting planetary and brown dwarf companions may change by factors of $\sim2$. Also the characteristics of the WDs most likely hosting a companion significantly depends on the local age-metallicity relation.

\subsection{Initial period distribution}

The perhaps most fundamental assumption in our population synthesis is that the companion occurrence rates derived by \citet{wolthoff2022} represent realistic approximations. 
Concerning the dependence of the occurrence rate on metallicity and stellar mass, this assumption appears to be well justified,
as complementary studies broadly agree with the results obtained by \citet{wolthoff2022}. 
For example, \citet{johnsonetal10-1} and \citet{fultonetal21-1} found a similar increase with metallicity and stellar mass up to $\sim2$\,\Msun,~and recent imaging surveys of FGK and A stars further confirm the trend that substellar objects are more frequent around more massive stars up to $1.5$\,\Msun\,\citep[e.g.,][]{viganetal17-1,wagneretal22-1}. 

With respect to the orbital period distribution of the companions, \citet{wolthoff2022} obtained good fits to the observational data using a broken power-law or log-normal distribution. In general, the companion occurrence rate increases with period up to a peak (turnover) and then decreases for longer periods. 
The general shape of the companion period distribution is in agreement with previous findings for giant planets around 
lower mass main-sequence stars. However, the peak (or turnover) 
found by \citet{wolthoff2022} is located at shorter periods (i.e., $718\,\pm\,226$ days for the broken power-law or $797\,\pm\,455$ days for the log-normal distribution) than derived from surveys of main-sequence stars
that found the peak to be close to the snow line \citep{Fernandesetat19}. 
This difference is somewhat surprising because, assuming the relationship between the peak of the occurrence rate and the snow line is a general trend, one would expect companions around more massive stars to be located farther away from the host star; instead, they seem to be found closer to it. 
Possible explanations for this discrepancy range from differences in the companion formation process, for example because the disk lifetimes are shorter around more massive stars \citep[][]{roncoetal24-1}, to the impact of stellar evolution (where tidal evolution may have already played a role) or false positives (see also \citealt{wolthoff2022} for a discussion).

Indeed, it has previously been argued that pulsations of very massive and luminous giants can mimic the radial-velocity signal produced by companions \citep{Saioetal-15}. In particular, \citet{Dollinger2021} identified several stellar properties associated with suspected radial velocity
planet candidates around evolved stars; i.e., large radii, low metallicities, low effective temperatures, and low surface gravities. In particular, giants with $R\gtrsim21\,R_\odot$ show an anomalously high incidence of radial velocity planet candidates, with orbital periods between $300$ and $800$\,days. These last two criteria would compromise the validity of at least three of the $37$ planets used by \citet{wolthoff2022}. 
In line with this, a recent study by \citet{Spaethetal25} demonstrates that one of these planet candidates (HIP\,16335b) is in fact a false positive, because the radial velocity oscillations can be explained by intrinsic stellar processes.  

If the period distribution by \citet{wolthoff2022} is indeed affected by a false positive or stellar evolution, our predictions would be directly affected, because we assumed their period distribution to generate a representative sample of potential WD progenitors with substellar companions. 
To explore how much a shift in the period distribution would affect our results, we moved the assumed period distribution to larger periods (simply by multiplying by a factor of $1500/800=1.875$) to produce a peak at $1500$\,days, which is consistent with \citet{Fernandesetat19}. The results of the simulations are displayed in rows $4$ and $5$ in Table\,\ref{tab:summary_results} and show an increase in the survival rate ($f_\mathrm{surv}$ increases by a factor of $2.2$ for weak tides and $3.1$ for strong tides). 
The final separation distribution follows the expected scaling, as the median increases by almost a factor of $(1500/800)^{2/3}$. 
(the median increase is slightly smaller, especially under \citetalias{Villaver_2009}, because closer companions are able to survive). 

In summary, as expected, the predicted companion occurrence rate depends on the assumed period distribution, which remains uncertain. However, even by moving the period distribution to wider semimajor axes, we estimate that less than 3\,\% of the current WD population hosts a substellar companion. 

\subsection{Eccentricity}

In our population synthesis, we assumed the eccentricity distribution derived by 
\citet{Kane24}. They used planets with masses exceeding that of Saturn from the NASA Exoplanet Archive \citep{Akesonetal13}. 
Most of the host stars with known companions and determined eccentricity are solar or lower mass stars, i.e., smaller than those of WD progenitors. 
Moreover, a recent study by \citet{Alqasimetal25} suggests 
that the planet eccentricity of transiting long-period giant planets depends on stellar metallicity, planet radius, and planet multiplicity. Such dependencies or a relation with any properties of the host star or the companions were not explored in detail by \citet{Kane24}. A more complete and detailed eccentricity distribution for giant planets around WD progenitors has yet to be derived. 

To very broadly explore the impact of the assumed eccentricity distribution, we simulated the survival of companions around WD progenitors assuming only circular orbits (sixth and seventh rows of Table \ref{tab:summary_results}). We found that the survival rate increases by more than a factor of $1.3$ for both tidal prescriptions (close to a factor o $1.4$ in the case of strong tides). Interestingly, under \citetalias{rasioetal96-1}, the median of the final semimajor axis increases. Although more companions closer to the star are able to survive, at the same time companions that previously survived (assuming the eccentricity distribution from \citealt{Kane24}) are less affected by tides, and so their final semimajor axis is larger. The characteristics of the companion-hosting WDs are very similar to those predicted by the corresponding baseline simulations.

\subsection{Impact of the mass-loss parameter}

Another crucial factor affecting the results is the modeling of stellar mass loss. For the MESA simulations, we adopted mass-loss prescriptions from \citet{Reimers1975}, with an efficiency of $\eta=0.5$ for the RGB, and the model from \citet{Bloecker95-1}, with $\eta=0.02$ for the AGB \citep{ventura-12}. Although these mass-loss prescriptions are widely accepted and in general observationally rather well constrained, the 
efficiency of mass loss on the AGB is still somewhat uncertain \citep[see][their introduction]{Cinquegranaetal-22}. 
In general, a smaller value of $\eta$ produces more thermal pulses and an extended AGB phase, while larger values truncate the late AGB phases much earlier. This clearly affects the survivability of companions. 

To analyze how this mass-loss efficiency on the AGB impacts our results, we ran our simulations adopting a significantly larger efficiency parameter ($\eta=0.7$) and found, as expected, that the survival rate increases, especially for stronger tides (see  bottom two rows in Table \ref{tab:summary_results}). Noticeably, the median of the final semimajor axis becomes considerably larger under stronger tides (compared to the baseline model), because the star expands and pulsates less, and the surviving companions are therefore more affected by mass loss and much less by tides. Even with this change, we still predict that under 3\,\% of the current WD population should host a substellar companion.

\subsection{The potential impact of additional objects}

One limitation in our predicted population is that we only considered single-companion systems; i.e., we ignored the potential impact of multiple companions. The related uncertainties are particularly important to discuss as 
multi-planetary
systems could be more common than single-planet systems \citep{Zhu2018,Zink2019}. For instance, \citet{Muldersetal-18} reproduced Kepler data with simulated planet populations and found that roughly two thirds of all stars could host multi-planetary systems. However, it remains unclear to what extent multiplicity affects our predictions. First, 
multi-planetary systems seem to be less frequent around more massive stars. Second,  
how multiple planets affect survival has not been investigated in detail. \citet{maldonadoetal21-1} and \citet{Maldonadoetal-22} combined stellar evolution and N-Body simulations, but investigated the evolution during the WD phase rather than the survival during the giant branches. They found that the higher the number of planets is, the more likely instabilities in the planetary system are during the WD phase. 

Furthermore, planets at low separations to WDs may result from a variety of dynamical processes involving the orbital perturbation induced by another companion, such as the eccentric Kozai-Lidov mechanism \citep{Stephan-21,O'Connor-21} 
or planet-planet scattering \citep{maldonadoetal21-1}. An additional companion located closer to the star might also truncate stellar evolution on the giant branches by generating common envelope evolution and thereby increase the chances of survival for
the more distant companions \citep[e.g.,][]{lagosetal21-1}. A systematic population study including all these effects, however, is currently out of reach.  

In general, it might also be that the survival of most companions would be little affected by multiplicity. \citet{roncoetal20-1} incorporated stellar tides, stellar evolution, and N-body simulations for two-planet systems. They find that the
interactions between planets changed the final fate (engulfment or survival)
of one of the companions in only roughly two per cent (2/99) of their simulations. This is because mean-motion resonances are required for this to happen.
We therefore believe that our predictions are valid first-order estimates of the population of substellar companions around WDs.

\subsection{Survival of engulfed companions}

In this work, we considered that all substellar companions that enter the giant envelope are destroyed, although this may not always be the case. In fact, massive planets around less massive stars have been suggested to survive the common-envelope phase. If orbital energy is efficiently used to unbind the envelope, engulfed companions may spiral in toward the common center of mass, but can avoid being evaporated or disrupted at the Roche limit \citep{Nelemans-98,Nordhaus-13}. However, this scenario most likely only applies to the most massive companions considered here. For instance, \citet{O'Connor-23} studied the potential survival of engulfed planets and found that for planets with masses of $1-10$\Mjup~around stars with masses of $1.0-1.5$\Msun,~survival was impossible. 

This situation might change if several planets are engulfed. In fact, common envelope evolution can explain the existence of substellar companions very close to WDs, especially when invoking successive planet engulfment or contributions from additional energy sources such as recombination energy \citep{lagosetal21-1,Chamandy-21,Merlov-21}.
We will investigate the number of companions that might survive engulfment in our population synthesis approach in a future work.

\subsection{Observability implications}

A key question that we did not answer in the present paper is how many of the predicted companions will be detectable with current and future observing facilities. A complete observational census of companions to WDs might allow us to constrain some of the uncertainties in our simulations, such as the strength of stellar tides, the mass-loss efficiency during the AGB phase, or maybe even the assumed initial companion distributions. 

The potential to directly detect the predicted companions through imaging will depend on the age of the WD, the total age of the system, and the orbital separation, which converts to the required contrast and angular resolution of the chosen observing facility. Detecting the companions through an infrared excess in the spectral energy distribution will depend on the relative brightness of the WD and the companion.  
We will determine the detectability of the predicted companions and present the results in a future publication.

\section{Conclusion}

We generated a representative population of potential WD progenitor stars with substellar companions using initial distributions based on the results of radial-velocity surveys of $1-4$\,\Msun~giant stars and a global age-metallicity relation. We then evolved these stars to the present day taking into account stellar and orbital evolution to 
predict the population of substellar companions around WDs. 
We find that under $3\pm1.5$\,\% of WDs are expected to host a substellar companion, suggesting that substellar companions around WDs are intrinsically rare. Roughly $\sim95$\,\% of the predicted substellar companions are gas giant planets. 
The exact fraction of WDs with companions depends on initial distributions, stellar tides, and mass loss during the AGB. 

To some extent, the properties of the surviving companion population are shaped by the adopted tidal prescription. 
The fraction of companions that survive the stellar evolution of their host star is roughly two times higher when adopting the weak-tides prescription of \citet{Villaver_2009} compared to the strong stellar tides suggested by \citet{rasioetal96-1}. 
Stronger tides generally require companions to be initially located at somewhat larger semimajor axes and to orbit less massive stars in order to survive, which results in a distribution shifted toward slightly larger final semimajor axes and lower WD masses. 
The companion population is expected to be located at semimajor axes from around $3$ to $24$\,au, with a median of $11$\,au, and is more likely to orbit lower mass white dwarfs  ($\sim 0.53M_\odot$ to $\sim 0.66M_\odot$) with relatively low total system ages ($\sim1$-$6$ Gyr), where the occurrence reaches values of $\gappr3$\,\%. 

Our general findings are largely independent of the assumed period and eccentricity distributions as well as the stellar mass-loss parameter. The local sample of white dwarfs close to the Sun might host more companions than we predict here, as the local age-metallicity relation implies a larger occurrence rate ($\lappr8$\,\%). 
In a future article, we aim to derive concrete predictions concerning the observability of the predicted companions in order to inform ongoing surveys of substellar companions to white dwarfs.

\begin{acknowledgements}
AMS, MRS, DC, JP, CRJ, JV, MPR thank for support from FONDECYT (grant No.\,1221059). MPR is partially supported by PICT-2021-I-INVI-00161 from ANPCyT, Argentina. FLV acknowledges support from ANID FONDECYT Postdoctoral Grant No. 3250464.
\end{acknowledgements}

\bibliographystyle{aa} 
\bibliography{WDplanets} 

\end{document}